\documentclass[final,3p,times,twocolumn]{elsarticle}

\usepackage{amssymb}
\usepackage{amsmath}
\usepackage{url}
\usepackage{graphicx}
\usepackage{subfig}
\usepackage{graphicx}
\usepackage{float}
\usepackage{rotating}
\usepackage{orcidlink}
\usepackage{tikz}
\usepackage{subcaption}
\usetikzlibrary{arrows.meta, positioning, calc}

\usepackage{hyperref}
\usepackage{enumitem}
\usepackage{stfloats}
\usepackage{amsthm}

\definecolor{orcidlogocol}{HTML}{A6CE39}

\newcommand{\orcidicon}{%
    \tikz[baseline=-0.5ex]\node[shape=circle,fill=orcidlogocol,inner sep=1pt] {\tiny\textsf{ID}};%
}

\newcommand{\orcid}[1]{\href{https://orcid.org/#1}{\orcidicon}}

\usepackage{booktabs}

\begin{document}

\begin{frontmatter}



\title{Scale-dependent temporal signatures of arboviral transmission in urban environments}


\author[1,2]{Marcílio Ferreira dos Santos}
\ead{marcilio.santos@ufpe.br}
\ead[url]{https://orcid.org/0000-0001-8724-0899}

\author[2,3]{Cleiton de Lima Ricardo}
\ead{cleiton.lricardo@ufpe.br}
\ead[url]{https://orcid.org/0000-0002-7114-1201}


\affiliation[1]{
    organization={Núcleo de Formação de Docentes, Universidade Federal de Pernambuco (UFPE)}, 
    addressline={}, 
    city={Caruaru},
    postcode={},
    state={PE},
    country={Brazil}
}

\affiliation[2]{
    organization={Núcleo Interdisciplinar de Ciências Exatas e da Natureza (NICEN), Universidade Federal de Pernambuco (UFPE)}, 
    addressline={}, 
    city={Caruaru},
    postcode={},
    state={PE},
    country={Brazil}
}

\begin{abstract}
Understanding epidemic dynamics in urban environments requires models capable of capturing interactions across space and time while accounting for biological constraints. In this work, we propose a probabilistic spatiotemporal framework based on pairwise interaction kernels to analyze the transmission dynamics of arboviruses using large-scale georeferenced data from Recife, Brazil. The model describes interactions between events as a function of spatial distance and temporally delayed influence, incorporating biologically motivated constraints such as finite interaction windows and incubation periods. Parameters are estimated via maximum likelihood, enabling the inference of interpretable quantities such as characteristic interaction time and effective spatial range. Our results reveal a marked asymmetry between spatial and temporal components. The spatial parameter systematically collapses, indicating that spatial proximity does not provide discriminatory information between diseases at the urban scale considered. In contrast, temporal dynamics exhibit a scale-dependent behavior: statistical differentiation between dengue, Zika, and chikungunya emerges only beyond a critical temporal window. We show that unconstrained models primarily capture short-term co-occurrence patterns, leading to apparent but non-robust differences between diseases. When biologically plausible constraints are imposed, these differences largely disappear, indicating that transmission dynamics share a common underlying structure. Additionally, the reconstruction of most-likely transmission networks reveals localized and structured interaction pathways consistent with plausible epidemic propagation mechanisms. These findings demonstrate that epidemic differentiation is not an intrinsic property observable at all scales, but rather an emergent phenomenon dependent on the temporal resolution of analysis. More broadly, the proposed framework highlights the importance of incorporating biological realism and scale-aware modeling in the interpretation of spatiotemporal epidemic data.
\end{abstract}
\begin{graphicalabstract}
\end{graphicalabstract}

\begin{highlights}

\item A probabilistic spatiotemporal framework is proposed to model arboviral transmission
\item Significant temporal differences are identified between dengue, Zika, and chikungunya
\item Spatial interaction does not discriminate between diseases in the urban setting
\item Temporal differentiation emerges only beyond a critical interaction window
\item Epidemic dynamics exhibit scale-dependent behavior across temporal horizons
\item A most-likely transmission network reveals structured pathways of interaction

\end{highlights}


\end{frontmatter}



\section{Introduction}

Understanding the dynamics of epidemic processes in urban environments remains a central challenge in complex systems and applied mathematics. The spread of infectious diseases is inherently driven by interactions occurring across space and time, giving rise to emergent patterns that reflect both biological mechanisms and underlying social and environmental structures \cite{keeling2008,chinazzi2020}. In densely populated urban settings, these interactions become particularly intricate, as mobility, population density, and environmental heterogeneity jointly shape transmission dynamics \cite{massaro2019,parselia2019}.

Arboviral diseases such as dengue, Zika, and chikungunya provide a compelling case for investigating these processes. Transmitted by a common vector, these diseases share similar spatial contexts but exhibit distinct epidemiological characteristics, including differences in incubation periods, infectiousness, and temporal evolution \cite{teixeira2009,honorio2009}. Moreover, the spatial distribution of these diseases is strongly influenced by socio-environmental factors, including urban infrastructure, climate variability, and vector ecology \cite{barcellos2001,borges2024}. This raises a fundamental question: to what extent can these diseases be distinguished based on their observed spatiotemporal dynamics?

Traditional approaches to epidemic modeling often rely on compartmental frameworks or spatial statistical methods. Compartmental models provide a mechanistic description of disease dynamics but typically neglect fine-grained spatial structure \cite{keeling2008}. Conversely, spatial and spatiotemporal statistical models emphasize geographic patterns but may treat interactions in an aggregated manner \cite{anselin1995,blangiardo2015}. While these approaches have been successful in many contexts, they may fail to capture the interaction-level structure underlying transmission events, particularly when high-resolution georeferenced data are available.

In recent years, there has been increasing interest in frameworks that model interactions directly, allowing the reconstruction of implicit transmission structures from observed data. These approaches are closely related to concepts from network science and complex systems, where interactions between entities give rise to emergent global behavior \cite{Estrada2012, EstradaHatano2008}. In particular, probabilistic interaction models provide a flexible way to describe how past events influence future occurrences without requiring explicit knowledge of transmission pathways.

In this work, we adopt a probabilistic perspective based on pairwise spatiotemporal interaction kernels. Instead of explicitly modeling transmission chains, we infer interaction patterns by assigning likelihoods to potential connections between events as a function of spatial distance and temporal delay. This approach provides an interpretable framework that captures essential features of interaction while avoiding strong assumptions about the underlying generative process.

Using a large-scale dataset of georeferenced arboviral cases from Recife, Brazil \cite{dataset_recife_2025}, we investigate whether distinct diseases exhibit identifiable signatures in their spatiotemporal dynamics. Our analysis reveals that apparent differences between diseases are strongly dependent on the temporal scale and modeling assumptions. In particular, we show that temporal parameters provide statistical differentiation only when short-term interactions are allowed, while spatial interaction appears to be largely homogeneous within the urban environment.

A key contribution of this work is the identification of scale-dependent behavior in epidemic differentiation. Through a sensitivity analysis on the temporal interaction window, we demonstrate that disease-specific signatures emerge only beyond a critical temporal scale. At shorter time horizons, the inferred dynamics become statistically indistinguishable across diseases, indicating a common underlying transmission structure. This finding highlights the importance of temporal scale as a fundamental dimension in the analysis of epidemic processes.

In addition, we reconstruct the most probable transmission network based on the inferred interaction probabilities, providing a network-based perspective on epidemic propagation. This complements the statistical analysis and reveals structured pathways of interaction that are consistent with localized transmission mechanisms.

Overall, our results suggest that arboviral dynamics in dense urban environments are best understood as a scale-dependent process, in which both homogeneity and differentiation emerge from the same system under different observational regimes. More broadly, the proposed framework demonstrates how simple probabilistic interaction models can reveal nontrivial structural properties of epidemic systems, including the emergence of scale-dependent behavior.
\section{Methodology}

\subsection{Data Description}

We consider a dataset of georeferenced arboviral cases (dengue, Zika, and chikungunya) collected in the city of Recife, Brazil. Each event is represented as a pair $(x_i, t_i)$, where $x_i \in \mathbb{R}^2$ denotes the spatial location in projected coordinates (UTM) and $t_i \in \mathbb{R}$ represents the time of occurrence.

All spatial data were transformed to a common coordinate reference system (SIRGAS 2000 / UTM zone 25S) to ensure consistency in distance calculations.

To control computational complexity and avoid bias from highly dense clusters, subsampling was applied when necessary, preserving the overall spatial and temporal structure of the dataset.

\subsection{Spatiotemporal Interaction Model}

We model the interaction between events using a pairwise spatiotemporal kernel. For each event $j$, the interaction intensity is defined as:

$$
\lambda_j = \sum_{i < j} f(d_{ij}, \Delta t_{ij}),
$$

where $d_{ij} = \|x_j - x_i\|$ is the spatial distance and $\Delta t_{ij} = t_j - t_i$ is the temporal delay.

The interaction kernel is defined as:

$$
f(d, \Delta t) = \exp(-\alpha d)\cdot g(\Delta t - \tau),
$$

where $\alpha > 0$ controls spatial decay and $\tau > 0$ represents a temporal delay introduced to enforce biological plausibility.

\subsection{Temporal Kernel}

The temporal component is modeled using a Gamma distribution:

$$
g(\Delta t) = \frac{\beta^k \Delta t^{k-1} e^{-\beta \Delta t}}{\Gamma(k)}, \quad \Delta t > 0,
$$

applied to the shifted variable $(\Delta t - \tau)$.

The parameters $\beta > 0$ and $k > 0$ control the rate and shape of the interaction profile, allowing flexibility in modeling both concentrated and dispersed temporal dynamics.

\subsection{Biological Constraints}

A key aspect of the model is the incorporation of biologically motivated constraints.

First, a temporal delay $\tau$ is introduced to account for intrinsic and extrinsic incubation periods in arboviral transmission. This prevents interactions at unrealistically short time intervals and distinguishes transmission from mere co-occurrence.

Second, interactions are restricted to a finite temporal window:

$$
0 < \Delta t \leq \Delta_{\max},
$$

where $\Delta_{\max}$ represents the maximum plausible delay between related events.

Third, a spatial cutoff is imposed:

$$
d \leq d_{\max},
$$

where $d_{\max}$ reflects the limited mobility of the mosquito vector and constrains interactions to biologically plausible distances.

Together, these constraints ensure that the inferred interaction structure remains consistent with known epidemiological mechanisms.

\subsection{Maximum Likelihood Estimation}

Model parameters $(\alpha, \beta, k)$ are estimated via maximum likelihood. The log-likelihood function is defined as:

$$
\mathcal{L} = \sum_{j} \log \lambda_j.
$$

To improve numerical stability, the temporal kernel is evaluated in log-space, and a regularization term is added to penalize extreme parameter values:

$$
\mathcal{L}_{\text{reg}} = \mathcal{L} - \gamma(\alpha^2 + \beta^2 + k^2),
$$

where $\gamma > 0$ is a small regularization coefficient.

Optimization is performed using bounded quasi-Newton methods, ensuring that parameters remain within physically meaningful ranges.

\subsection{Computational Considerations}

To improve efficiency, spatial queries are performed using a k-d tree structure, restricting candidate interactions to a local neighborhood defined by $d_{\max}$.

Distances are rescaled to improve parameter identifiability, particularly for the spatial parameter $\alpha$, which is sensitive to the scale of measurement.

These optimizations allow the model to be applied to large datasets while maintaining numerical stability.

\subsection{Transmission Network Reconstruction}

Based on the inferred interaction kernel, we reconstruct a probabilistic transmission network. For each event $j$, the most likely parent event $i$ is selected according to:

$$
P(i \to j) = \frac{f(d_{ij}, \Delta t_{ij})}{\sum_{k<j} f(d_{kj}, \Delta t_{kj})}.
$$

This yields a directed graph representing the most probable interaction pathways between events, providing a network-based interpretation of epidemic propagation.

\subsection{Sensitivity Analysis}

To evaluate robustness, we perform sensitivity analyses with respect to:

\begin{itemize}
\item the temporal interaction window $\Delta_{\max}$
\item the spatial cutoff $d_{\max}$
\item the temporal delay parameter $\tau$
\end{itemize}

These analyses allow the identification of scale-dependent behavior and ensure that the observed patterns are not artifacts of specific parameter choices.

\subsection{Hawkes Process Extension}

As an extension, we consider a spatiotemporal Hawkes process:

$$
\lambda(x,t) = \mu + \sum_{t_i < t} K \cdot \exp(-\alpha d_i)\cdot g(t - t_i),
$$

where $\mu$ represents a background rate and $K$ controls the strength of triggered interactions.

Although the estimation is approximate, this formulation provides a more complete generative interpretation and allows comparison with established point process models.
\section{Probabilistic Framework for Spatiotemporal Interaction}

The analysis of arboviral transmission requires models capable of capturing both spatial proximity and temporal ordering of events. In this work, we adopt a probabilistic pairwise interaction framework that describes the influence of past cases on subsequent occurrences through a spatiotemporal kernel.

Let $\mathcal{D} = \{(x_i, t_i)\}_{i=1}^n$ denote the set of reported cases, where $x_i \in \mathbb{R}^2$ represents the spatial location of event $i$ and $t_i \in \mathbb{R}$ its time of occurrence. Without loss of generality, we assume the events are ordered such that $t_1 \leq t_2 \leq \cdots \leq t_n$.

\subsection{Pairwise Interaction Structure}

For each event $j$, we define an interaction intensity $\lambda_j$ that aggregates the influence of all preceding events:

$$
\lambda_j = \sum_{i < j} f(d_{ij}, \Delta t_{ij}),
$$

where $d_{ij} = \|x_j - x_i\|$ denotes the spatial distance between events and $\Delta t_{ij} = t_j - t_i$ the temporal lag.

This formulation assumes that each past event contributes additively to the occurrence of future events, without imposing a specific generative mechanism such as branching or background processes. Instead, it provides a flexible description of interaction structure that can be interpreted probabilistically.

\subsection{Spatiotemporal Kernel}

The interaction kernel is defined as a separable function:

$$
f(d, \Delta t) = \exp(-\alpha d)\, g(\Delta t - \tau),
$$

where $\alpha > 0$ controls spatial decay, and $\tau > 0$ represents a temporal delay.

The spatial component $\exp(-\alpha d)$ models the decay of interaction with distance. Larger values of $\alpha$ correspond to strongly localized interactions, while $\alpha \to 0$ indicates weak or negligible spatial dependence.

The temporal component is modeled using a shifted Gamma distribution:

$$
g(\Delta t) = \frac{\beta^k (\Delta t)^{k-1} e^{-\beta \Delta t}}{\Gamma(k)}, \quad \text{for } \Delta t > 0,
$$

which is applied to the shifted variable $(\Delta t - \tau)$. The parameters $\beta > 0$ and $k > 0$ control the rate and shape of the temporal interaction profile, respectively.

\subsection{Biological Interpretation of Temporal Delay}

The inclusion of a temporal delay $\tau$ is motivated by biological considerations. Arboviral transmission involves both intrinsic incubation in the human host and extrinsic incubation in the mosquito vector. As a consequence, transmission cannot occur instantaneously after infection.

By introducing $\tau$, we exclude interactions occurring at unrealistically short time intervals, ensuring that the inferred interaction structure reflects plausible transmission pathways rather than mere co-occurrence.

This modification plays a crucial role in distinguishing between aggregated epidemic intensity and true transmission dynamics.

\subsection{Probabilistic Interpretation}

The interaction intensity $\lambda_j$ can be used to define a probability distribution over potential parent events:

$$
P(i \to j) = \frac{f(d_{ij}, \Delta t_{ij})}{\lambda_j}.
$$

This expression represents the probability that event $i$ contributed to event $j$, given the entire history of prior events.

This formulation induces a weighted, directed network over the set of cases, where edges represent probabilistic transmission pathways. Importantly, this interpretation does not assume that transmission is uniquely determined, but rather that multiple past events may contribute with different weights.

\subsection{Interpretable Quantities}

The parameters of the model admit direct epidemiological interpretation.

The expected temporal delay between interacting events is given by:

$$
\mathbb{E}[\Delta t] = \frac{k}{\beta},
$$

which defines a characteristic time scale of interaction.

Similarly, the spatial parameter $\alpha$ defines an effective interaction radius:

$$
d_{\text{eff}} = \frac{1}{\alpha},
$$

representing the distance at which the interaction decays to a fraction $e^{-1}$ of its maximum.

These quantities provide a compact summary of the spatiotemporal dynamics and enable comparison across diseases.

\subsection{Modeling Perspective}

The proposed framework occupies an intermediate position between descriptive statistical models and fully generative processes such as Hawkes models. While it does not explicitly model background rates or branching structure, it retains a probabilistic interpretation and captures essential features of spatiotemporal interaction.

This simplicity allows for robust parameter estimation and facilitates the exploration of structural properties of the data, particularly in the presence of incomplete or noisy observations.

At the same time, the framework is sufficiently flexible to incorporate biological constraints and to reveal scale-dependent behavior, which plays a central role in the analysis presented in the following sections.
\section{Results and Interpretation of Spatiotemporal Dynamics}

\subsection{Spatial Distribution of Cases}

\begin{figure*}
\centering
\includegraphics[width=0.32\linewidth]{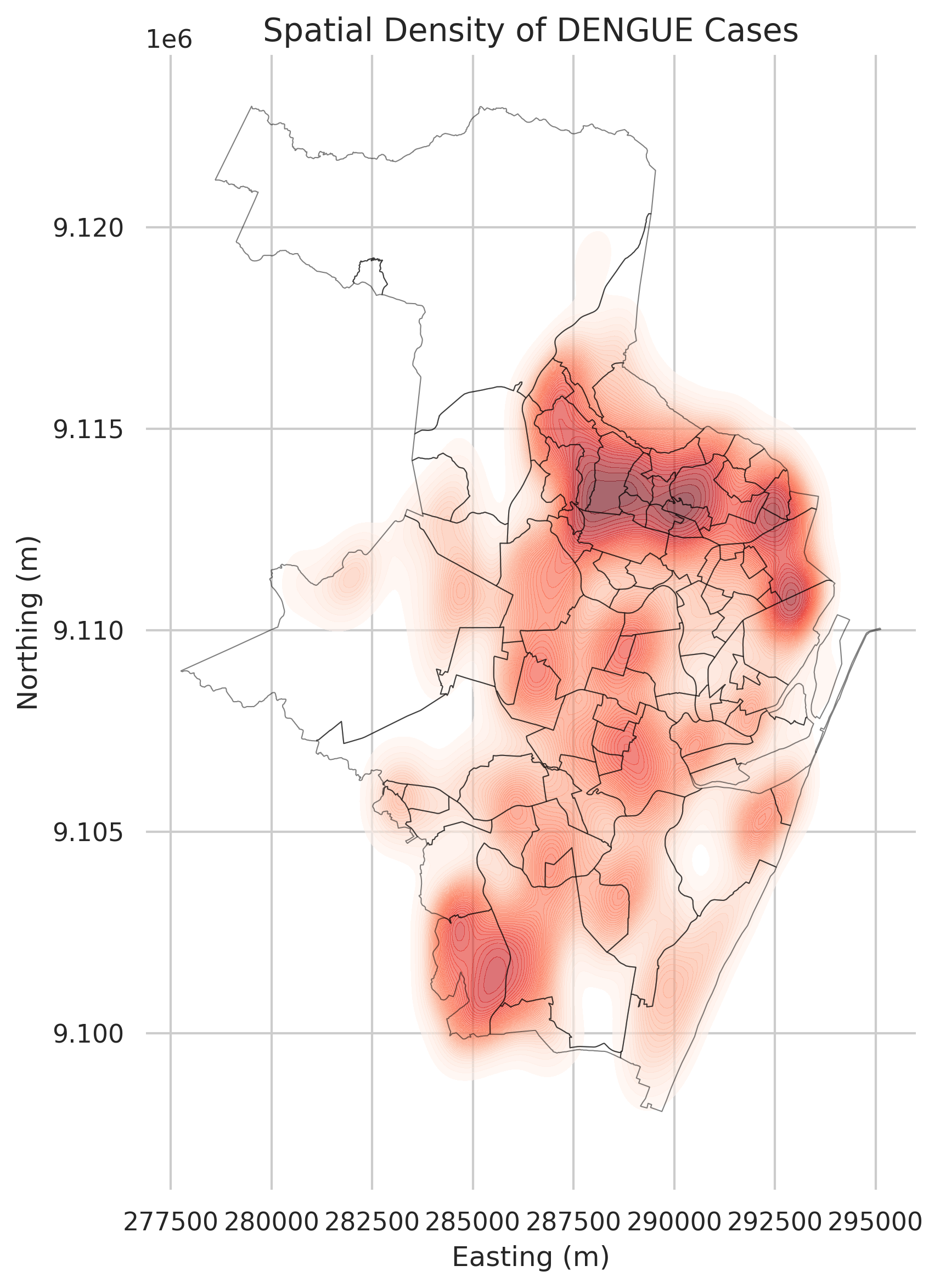}
\includegraphics[width=0.32\linewidth]{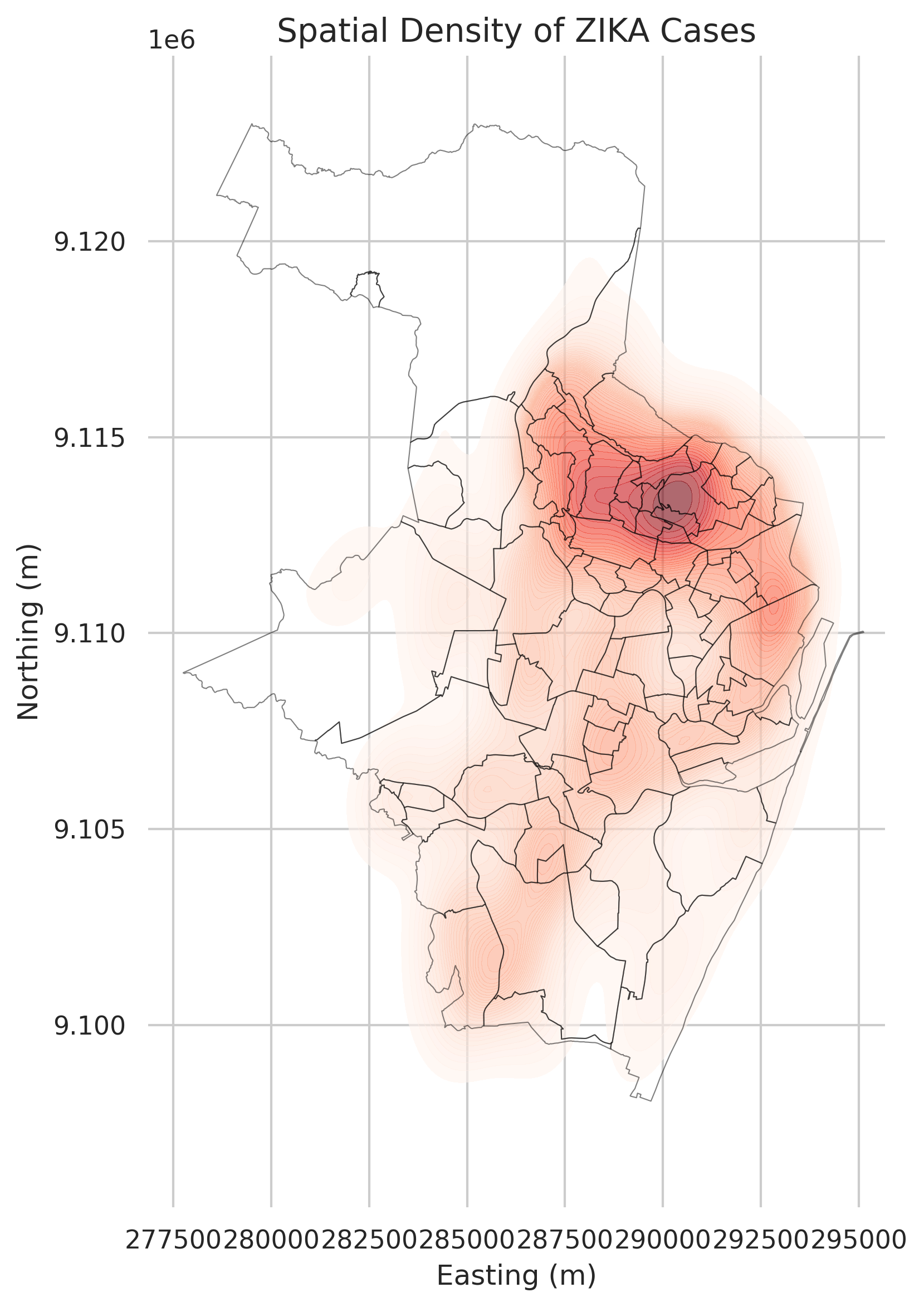}
\includegraphics[width=0.32\linewidth]{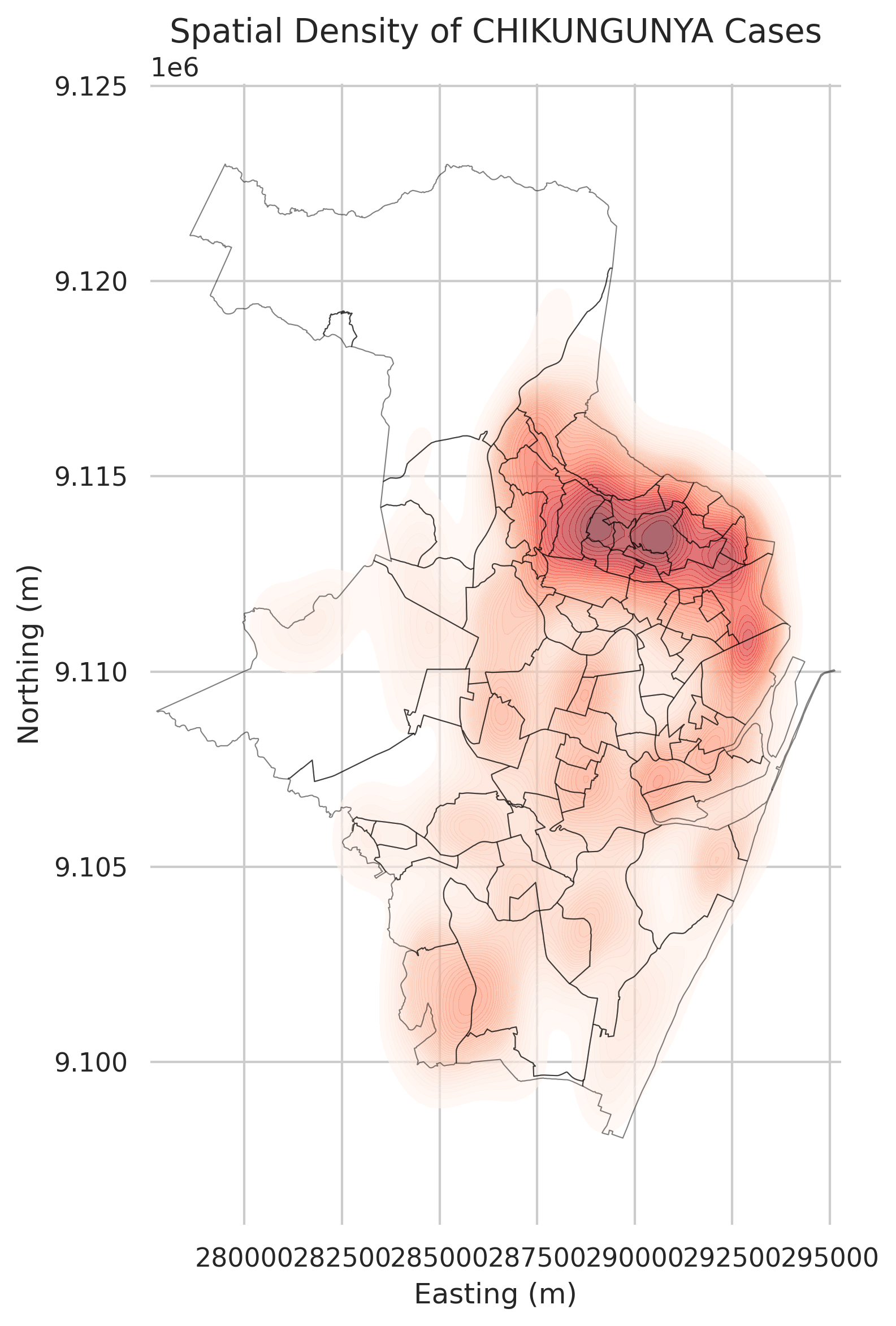}
\caption{Kernel density estimation of arboviral cases in Recife. The spatial distributions exhibit similar clustering patterns across diseases, with high-density regions concentrated in central and eastern urban areas.}
\label{fig:kde_maps}
\end{figure*}

We begin by examining the spatial distribution of arboviral cases in Recife. Figure~\ref{fig:kde_maps} shows kernel density estimates for dengue, Zika, and chikungunya.

The spatial distributions reveal pronounced clustering, particularly in densely populated urban regions. However, despite local variations in intensity, the overall structure of spatial concentration is remarkably similar across diseases.

This suggests that the spatial organization of cases is largely governed by shared environmental, demographic, and infrastructural factors, rather than disease-specific transmission mechanisms.

This observation provides an initial indication of spatial homogeneity, which is later confirmed by the collapse of the spatial parameter in the interaction model.
\subsection{Estimated Parameters and Global Behavior}

The estimated parameters under the biologically constrained model are summarized in Table~\ref{tab:parameters_summary}. Across all diseases, the parameters exhibit a high degree of consistency.

\begin{table}[H]
\centering
\caption{Estimated parameters under biologically constrained model}
\label{tab:parameters_summary}
\begin{tabular}{lccc}
\toprule
Disease & $\alpha$ & $\beta$ & $k$ \\
\midrule
Chikungunya & 0.001 & 0.621 & 2.87 \\
Dengue      & 0.001 & 0.626 & 2.93 \\
Zika        & 0.001 & 0.592 & 2.85 \\
\bottomrule
\end{tabular}
\end{table}

This convergence suggests that the three arboviruses share a common interaction structure when biologically plausible constraints are imposed.

\subsection{Absence of Spatial Differentiation}

A central result of this study is the systematic collapse of the spatial parameter $\alpha$. Across all configurations, $\alpha$ converges to its lower bound, indicating that spatial decay does not influence the interaction kernel.

This implies that spatial proximity does not provide discriminatory information between diseases at the urban scale considered. The observed spatial clustering in Figure~\ref{fig:map_points} therefore reflects a shared spatial substrate rather than disease-specific dynamics.

\subsection{Temporal Dynamics Under Constraints}

Temporal dynamics exhibit a markedly different behavior. In unconstrained models, differences between diseases emerge due to interactions at very small time intervals.

\begin{figure*}
\centering
\includegraphics[width=0.45\linewidth]{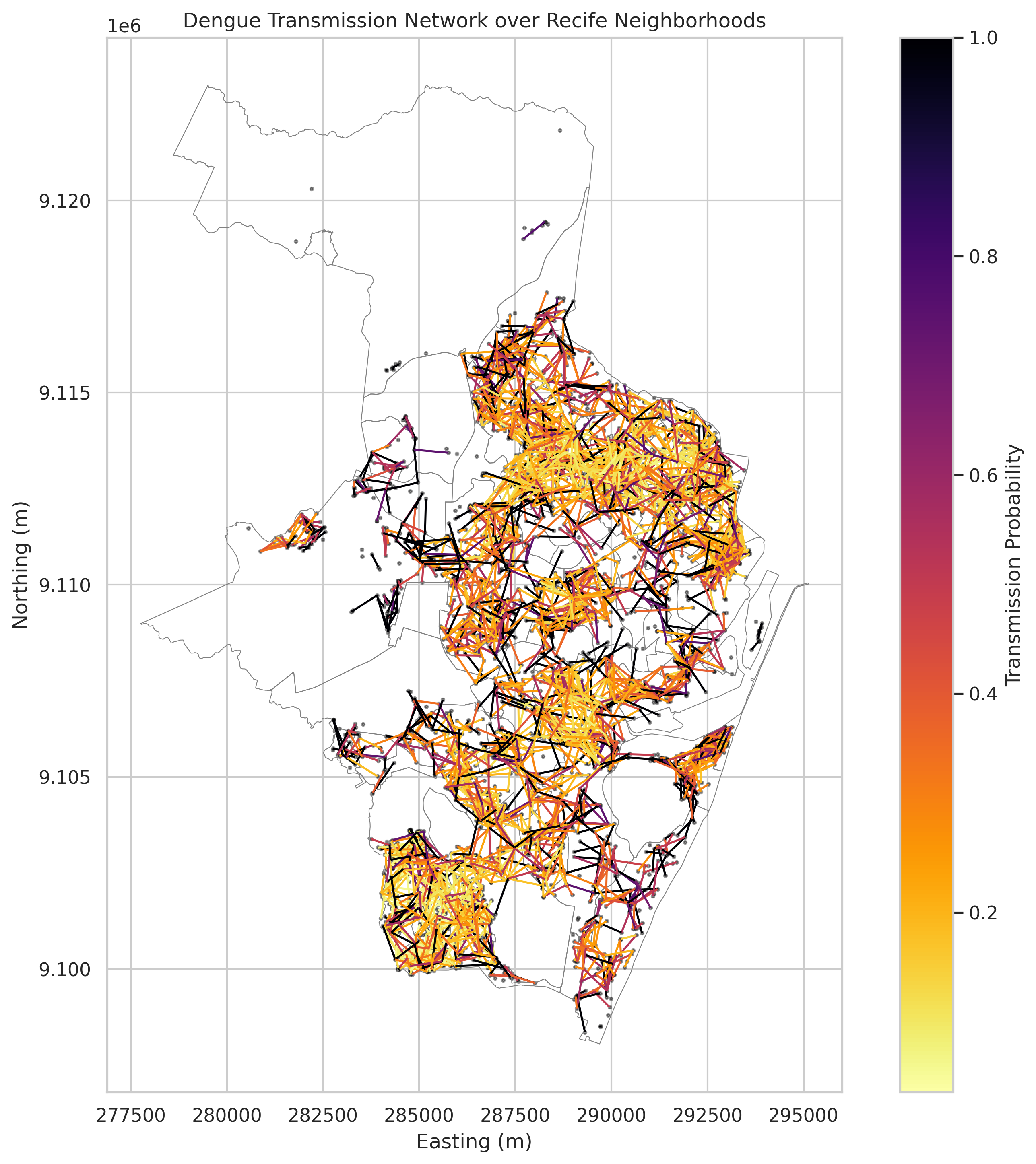}
\includegraphics[width=0.45\linewidth]{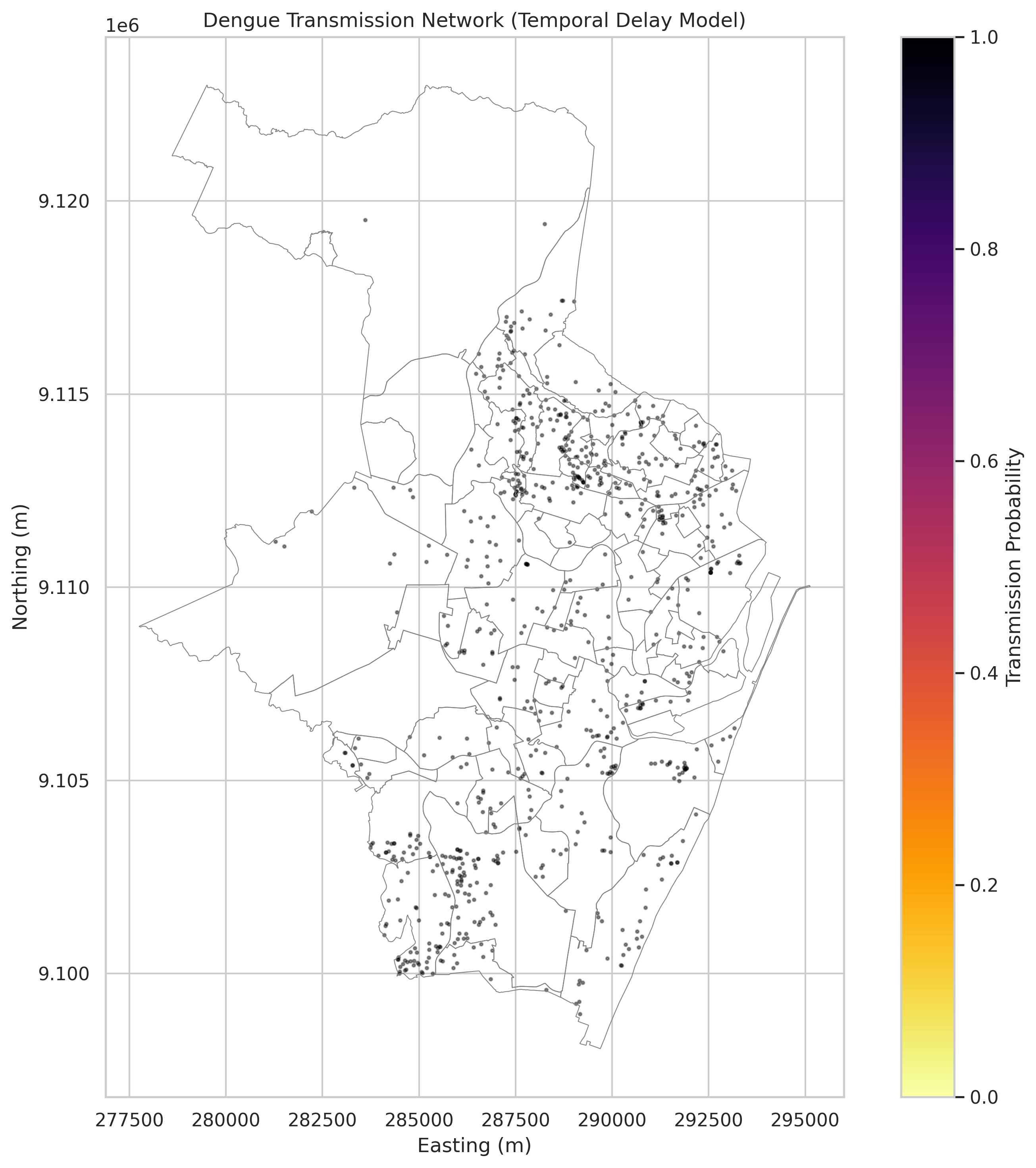}
\caption{Comparison between unconstrained (left) and biologically constrained (right) transmission networks.}
\label{fig:comparison_delay}
\end{figure*}

This effect is illustrated in Figure~\ref{fig:comparison_delay}, which compares the inferred transmission networks with and without temporal delay.

The unconstrained model produces dense and highly connected networks, reflecting short-term co-occurrence. In contrast, the constrained model yields more localized and structured interactions, consistent with plausible transmission dynamics.

\subsection{Structure of the Transmission Network}

\begin{figure}[H]
\centering
\includegraphics[width=0.8\linewidth]{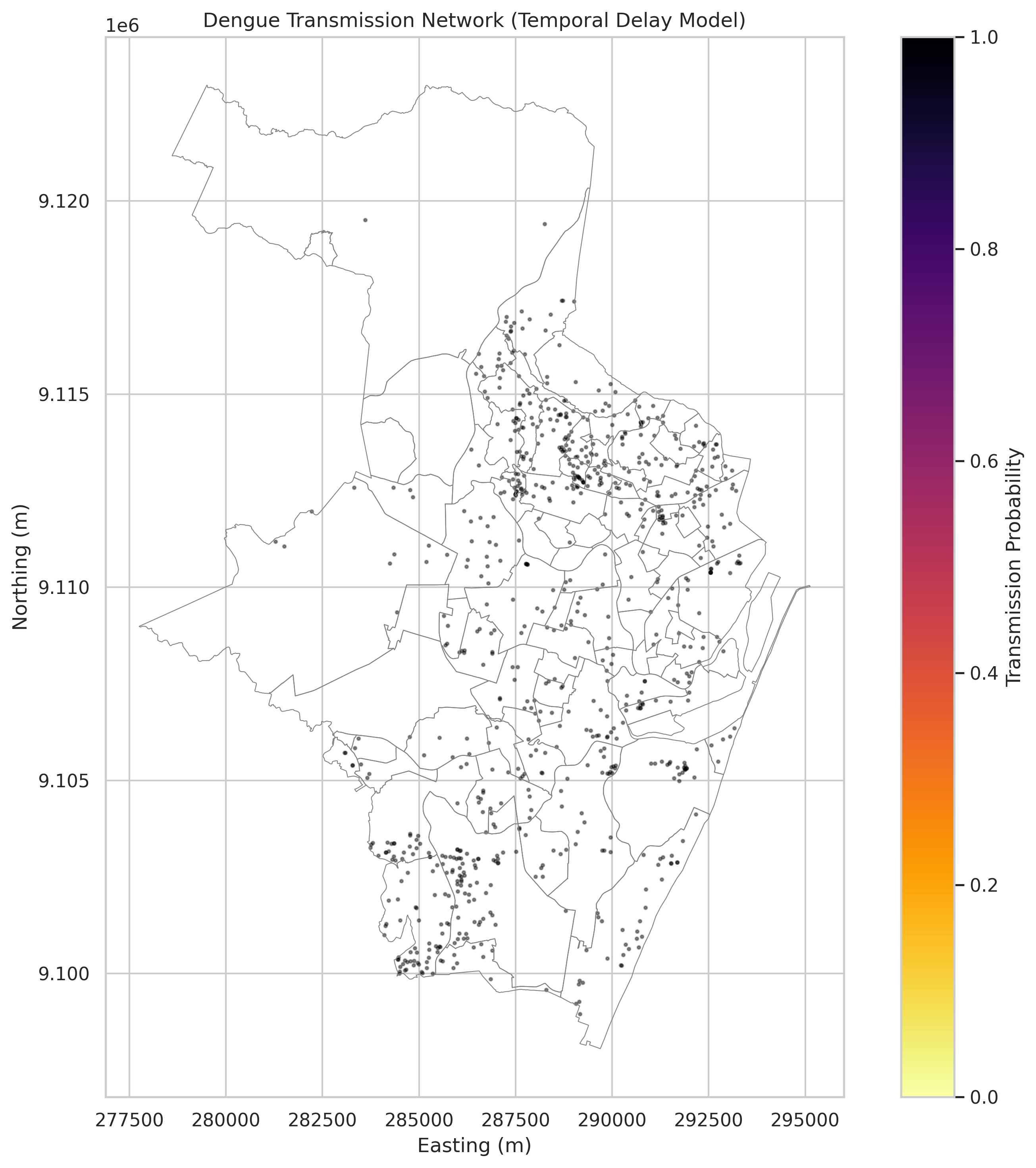}
\caption{Most probable transmission network inferred under biologically constrained model. Edge colors represent transmission probabilities.}
\label{fig:network}
\end{figure}

The transmission network inferred under biologically constrained conditions is shown in Figure~\ref{fig:network}.

The network exhibits localized connectivity patterns, with interactions confined to plausible spatial and temporal neighborhoods. Notably, long-range connections are suppressed, even in the absence of strong spatial decay at the parameter level.

This indicates that meaningful transmission structure emerges from the combination of temporal ordering and spatial constraints, rather than from explicit spatial decay.

\subsection{Distribution of Temporal Parameters}

\begin{figure*}
\centering
\includegraphics[width=1\linewidth]{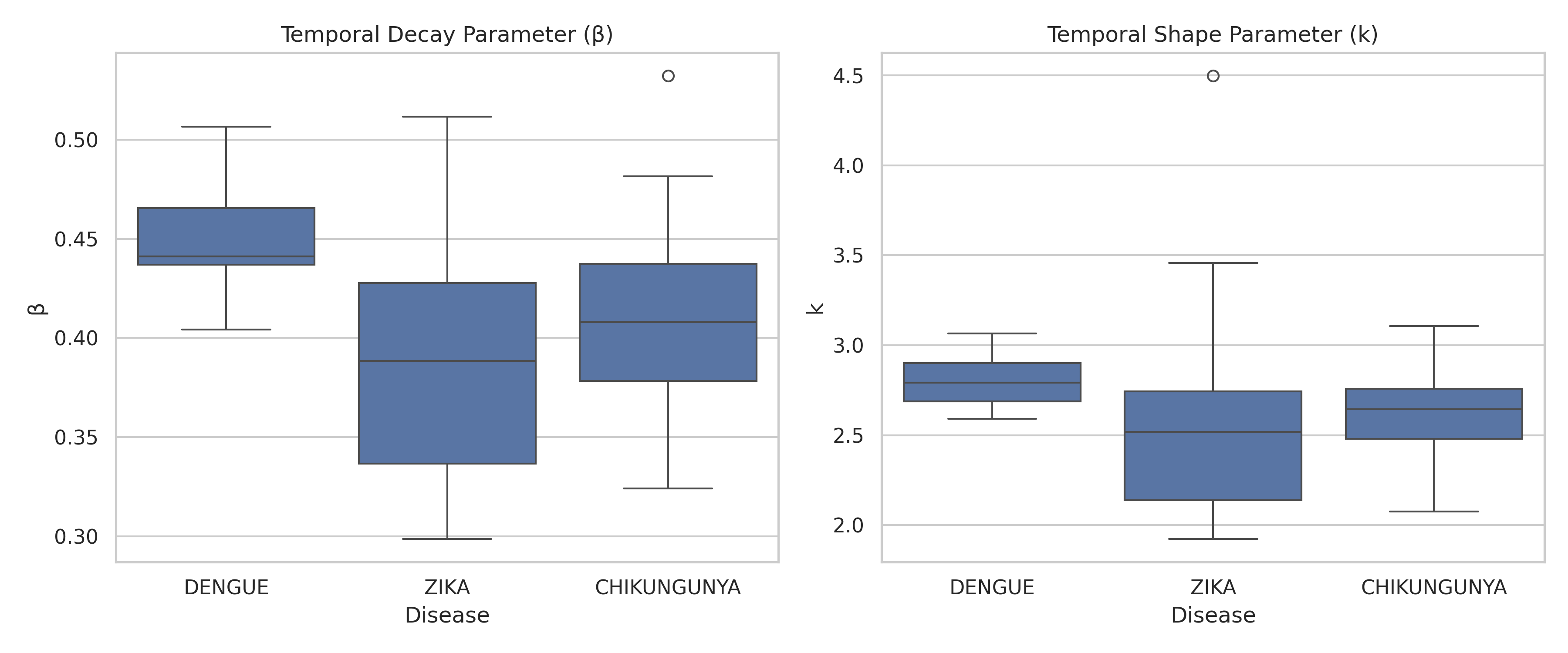}
\caption{Distribution of temporal parameters across diseases.}
\label{fig:boxplots}
\end{figure*}

The distribution of temporal parameters across diseases is shown in Figure~\ref{fig:boxplots}.

Under unconstrained conditions, differences between diseases are apparent. However, these differences diminish significantly when biologically plausible delays are introduced, leading to convergence of the parameter distributions.

\subsection{Emergence of Scale-Dependent Behavior}

To assess the role of temporal scale, we analyze the dependence of statistical differentiation on the interaction window.

\begin{table}[H]
\centering
\caption{Kruskal--Wallis test results}
\begin{tabular}{ccc}
\toprule
Window & $\beta$ & $k$ \\
\midrule
5  & 0.54 & 0.48 \\
7  & 0.23 & 0.16 \\
10 & 0.018 & 0.22 \\
\bottomrule
\end{tabular}
\end{table}

\begin{figure}[H]
\centering
\includegraphics[width=0.7\linewidth]{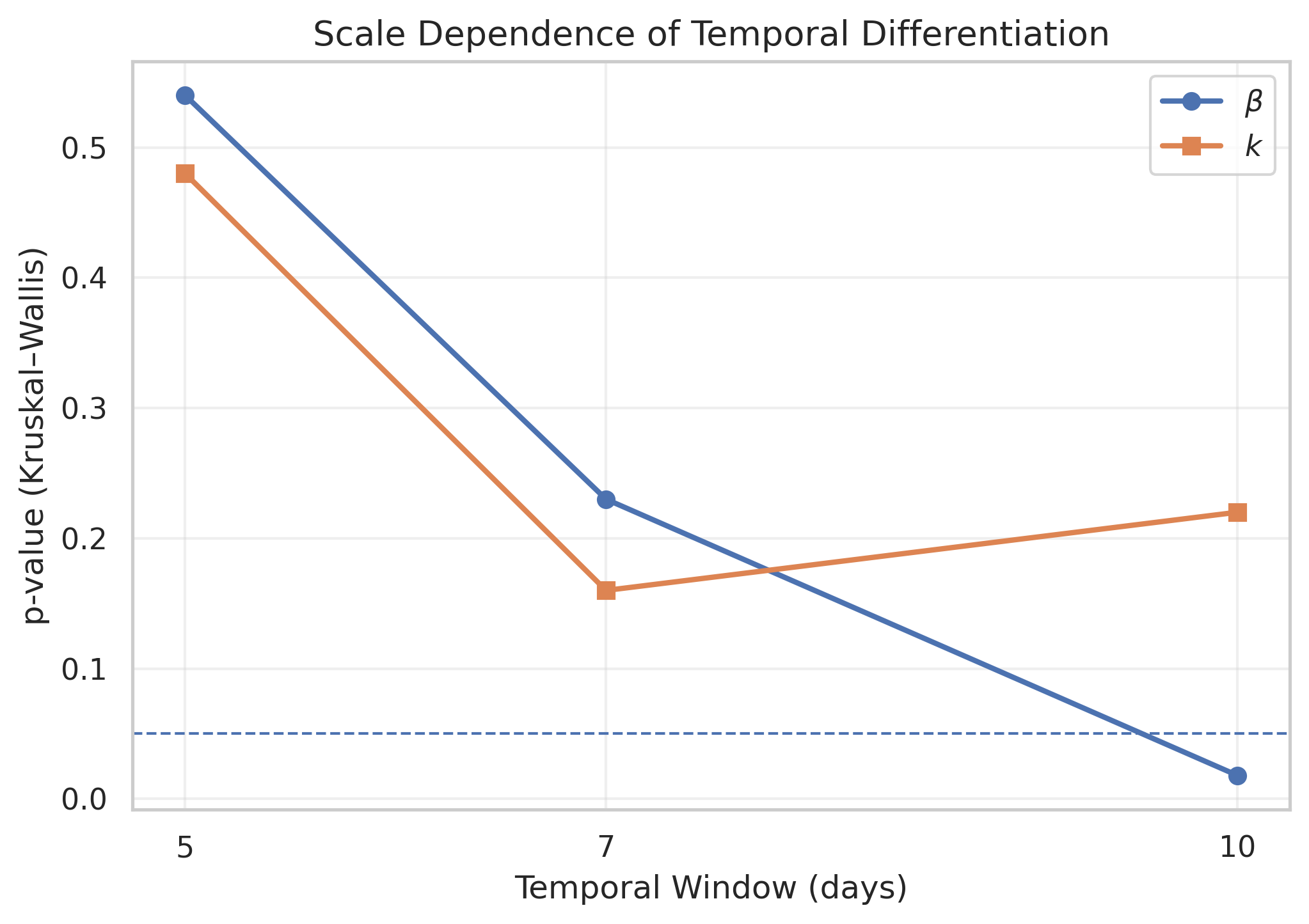}
\caption{Dependence of statistical differentiation on temporal window.}
\label{fig:scale}
\end{figure}

These results are further illustrated in Figure~\ref{fig:scale}.

At short temporal scales, the system behaves homogeneously, with no significant differences detected between diseases. As the temporal window increases, differences begin to emerge.

This behavior indicates the existence of a critical temporal scale beyond which disease-specific patterns become observable.

\subsection{Co-occurrence versus Transmission Regimes}

The combined analysis reveals two distinct regimes of interaction.

Without temporal delay, the model captures co-occurrence patterns, reflecting aggregated epidemic intensity. These patterns generate apparent differences between diseases.

With delay, the model isolates transmission dynamics, leading to convergence of parameters across diseases.

This demonstrates that co-occurrence and transmission correspond to fundamentally different regimes of the same process, and that distinguishing between them is essential for accurate interpretation.

\section{Discussion}

The results obtained in this study reveal a consistent and structurally meaningful pattern in the spatiotemporal dynamics of arboviral transmission in an urban environment. In particular, the analysis demonstrates that the interpretation of epidemic dynamics depends critically on the temporal scale and on the modeling constraints imposed.

Initial analyses based on unconstrained pairwise interaction models suggest that temporal parameters ($\beta$ and $k$) provide statistical differentiation between dengue, Zika, and chikungunya under certain temporal windows. However, this apparent differentiation is not robust. It arises primarily from interactions at very small time differences, which capture short-term co-occurrence rather than biologically plausible transmission events.

At the same time, the spatial parameter $\alpha$ systematically converges to values close to zero across all configurations, including unconstrained and constrained models. This behavior indicates a degeneracy in the spatial component of the kernel, effectively rendering the interaction process insensitive to distance. As a consequence, the unconstrained model may assign high interaction probabilities to events that are temporally close but spatially distant, reducing the inferred structure to a predominantly temporal process.

This finding highlights a fundamental limitation of unconstrained spatiotemporal kernels: without appropriate restrictions, temporal proximity dominates interaction inference, masking spatial structure and leading to a loss of identifiability in the spatial dimension.

To address this issue, we introduced biologically motivated constraints, including a spatial cutoff and a temporal delay. The spatial cutoff restricts interactions to distances compatible with the mobility of the mosquito vector, preventing unrealistic long-range connections. The temporal delay excludes interactions occurring at unrealistically short time intervals, ensuring that the inferred structure reflects plausible transmission pathways.

Under these constraints, a significant change in behavior is observed. The temporal parameters converge across diseases, and the previously observed differences largely disappear. This indicates that the differentiation seen in unconstrained models is driven by short-term co-occurrence patterns rather than intrinsic transmission mechanisms.

The sensitivity analysis further reveals that the ability to distinguish between diseases depends strongly on the temporal scale of observation. For short temporal windows, the dynamics appear statistically indistinguishable across diseases, suggesting a common underlying interaction structure. As the temporal window increases, statistically significant differences begin to emerge, indicating that aggregated effects introduce apparent differentiation at larger scales.

This scale-dependent behavior constitutes a key result of the study. It suggests that epidemic differentiation is not an intrinsic property observable at all resolutions, but rather an emergent phenomenon that depends on the temporal horizon considered. Such behavior is consistent with observations in complex systems, where macroscopic structure arises from the accumulation of local interactions across multiple scales.

The reconstruction of the most likely transmission network provides additional insight into the organization of epidemic dynamics. Under biologically constrained conditions, the inferred networks exhibit localized and structured connectivity patterns consistent with plausible propagation mechanisms. However, the absence of strong spatial differentiation at the parameter level indicates that these patterns are largely shared across diseases, reinforcing the conclusion that variability arises primarily from temporal and scale-dependent effects.

Finally, the extension toward a full spatiotemporal Hawkes process suggests that incorporating a background rate and explicit triggering mechanism may provide a more complete generative description of the data. Although the estimation remains approximate, the results indicate that the inclusion of biological and spatial constraints improves the stability of inference and helps reveal the underlying structure of interactions.

\section{Conclusion}

In this work, we proposed a probabilistic framework for analyzing the spatiotemporal dynamics of arboviral transmission using georeferenced data.

A central result of the study is the identification of strong scale-dependent behavior. We show that apparent differences between diseases emerge only under specific modeling conditions, particularly when short-term interactions are allowed. Once biologically plausible constraints are imposed, both spatial and temporal parameters converge, and the dynamics become statistically indistinguishable.

The spatial parameter consistently collapses to its lower bound across all configurations, indicating that spatial decay does not provide meaningful discriminatory information at the urban scale considered. This suggests that spatial structure is largely homogeneous across diseases and reflects shared environmental and infrastructural conditions.

The convergence of temporal parameters under constrained models indicates that arboviral transmission dynamics share a common underlying structure. The differences observed in unconstrained models are therefore better interpreted as manifestations of co-occurrence patterns and aggregated epidemic intensity, rather than intrinsic transmission mechanisms.

These findings highlight the importance of incorporating biological realism and multi-scale analysis in epidemic modeling. Without appropriate constraints, models may capture observational artifacts rather than true transmission dynamics.

Overall, this study demonstrates that arboviral dynamics in urban environments are best understood as a scale-dependent process, in which both homogeneity and differentiation emerge from the same system under different observational regimes.

\end{document}